\RequirePackage{arydshln}
\documentclass[aps,twocolumn,nofootinbib,superscriptaddress,preprintnumbers,pra,10pt]{revtex4-1}

\usepackage{float}
\usepackage{colortbl}
\usepackage{amsmath,amssymb}
\usepackage{dsfont} 
\usepackage{hyperref}
\usepackage{graphicx}
\usepackage{enumitem}
\usepackage{arydshln}
\usepackage{mathtools}
\usepackage{bbold}

\makeatletter
\g@addto@macro\bfseries{\boldmath}
\makeatother

\newcommand{\s}[1]{{\mbox{\tiny ($#1$)}}}
\newcommand{\be} {\begin{equation}}
\newcommand{\ee} {\end{equation}}
\newcommand{\bea} {\begin{eqnarray}}
\newcommand{\eea} {\end{eqnarray}}
\newcommand{\no} {\nonumber}
\newcommand{\gcav}{g_c^\prime}
\newcommand{\gBav}{g_Z^\prime}
\newcommand{\gBp}{g_B^\prime}
\newcommand{\gcl}{g_c^\s{{l}}}
\newcommand{\COLL}{C_\Omega}

\usepackage[usenames,dvipsnames]{xcolor}

\begin{document}

\preprint{ZU-TH-36/17}

\title{A three-site gauge model for flavor hierarchies and flavor anomalies}
 
\author{Marzia Bordone}
\email{mbordone@physik.uzh.ch}
\affiliation{Physik-Institut, Universit\"at Zu\"rich, CH-8057 Z\"urich, Switzerland}
\author{Claudia Cornella}
\email{claudia.cornella@physik.uzh.ch}
\affiliation{Physik-Institut, Universit\"at Zu\"rich, CH-8057 Z\"urich, Switzerland}
\author{Javier Fuentes-Mart\'{\i}n}
\email{fuentes@physik.uzh.ch}
\affiliation{Physik-Institut, Universit\"at Zu\"rich, CH-8057 Z\"urich, Switzerland}
\author{Gino Isidori}
\email{gino.isidori@physik.uzh.ch}
\affiliation{Physik-Institut, Universit\"at Zu\"rich, CH-8057 Z\"urich, Switzerland}

\begin{abstract}
\vspace{5mm}
We present a three-site Pati-Salam gauge model able to explain the Standard Model flavor hierarchies while, at the same time, accommodating the recent experimental hints of lepton-flavor non-universality in $B$ decays. 
The model is consistent with low- and high-energy bounds, and predicts a rich spectrum of new states at 
the TeV scale that could be probed in the near future by the high-$p_T$ experiments at the LHC.
\vspace{3mm}
\end{abstract}

\maketitle

\allowdisplaybreaks

\section{Introduction}\label{sec:intro}

Recent data on semileptonic $B$ decays  indicate 
anomalous violations of Lepton Flavor Universality (LFU) 
of short-distance origin. 
The statistical significance of each anomaly does not exceed the $3\sigma$ level,  
but  the overall set of deviations from the Standard Model (SM) predictions 
is very consistent. 
The evidences collected so far can naturally be grouped into two categories,
according to the underlying quark-level transition: i)~deviations from 
$\tau/\mu$ (and $\tau/e$) universality in  $b\to c \ell \bar\nu$
charged currents~\cite{Lees:2013uzd,Aaij:2015yra,Hirose:2016wfn,Aaij:2017deq};
ii)~deviations from $\mu/e$ universality in  $b\to s \ell \bar{\ell}$ neutral currents~\cite{Aaij:2014ora,Aaij:2017vbb}.
The latter turn out to be consistent~\cite{Altmannshofer:2015sma,Descotes-Genon:2015uva}
with the anomalies reported in the angular distributions of the $B^0 \to K^{*0} \mu^+ \mu^-$ decay \cite{Aaij:2013qta,Aaij:2015oid}.

A  common origin of the two set of anomalies is not obvious, but is very appealing from the theoretical point of view.
Severals attempts to provide a combined explanation  of the two effects have been presented in the recent literature~\cite{Bhattacharya:2014wla, Alonso:2015sja, Greljo:2015mma, Calibbi:2015kma,Bauer:2015knc,Fajfer:2015ycq, Barbieri:2015yvd, Das:2016vkr, Hiller:2016kry, Bhattacharya:2016mcc,Boucenna:2016wpr,Buttazzo:2016kid,Boucenna:2016qad,Barbieri:2016las,Bordone:2017anc,Crivellin:2017zlb,Becirevic:2016oho,Cai:2017wry,Megias:2017ove}.
Among them, a class of particularly motivated models 
are those based  on TeV-scale new physics (NP) coupled mainly to the third generation of SM fermions, 
with subleading effects on the light generations controlled by an approximate 
 $\mathrm{U(2)_Q}\times \mathrm{U(2)_L}$  flavor symmetry~\cite{Barbieri:2011ci}.
As recently shown in~\cite{Buttazzo:2017ixm}  (see also \cite{Greljo:2015mma,Barbieri:2015yvd,Bordone:2017anc}),
an Effective Field Theory (EFT) based on this flavor symmetry allows us to account for the observed semileptonic 
LFU anomalies taking into account the tight constraints from other 
low-energy data~\cite{Feruglio:2017rjo,Feruglio:2016gvd}. Moreover, the EFT fit singles out  the case of a
vector leptoquark (LQ) field $U_\mu \sim (\mathbf{3},\mathbf{1})_{2/3}$, originally proposed in~\cite{Barbieri:2015yvd},
 as the simplest and most successful framework with a single TeV-scale mediator
 (taking into account also the direct bounds from high-energy searches~\cite{Faroughy:2016osc}).
 
 While the results of Ref.~\cite{Buttazzo:2017ixm} are quite encouraging, the EFT solution and the simplified models
require an appropriate UV completion. In particular, the vector LQ mediator
 could be a composite state of a new strongly interacting sector, as proposed in \cite{Barbieri:2015yvd,Barbieri:2016las}, 
 or a massive gauge boson of a spontaneously broken  gauge theory, 
as proposed in \cite{Assad:2017iib,DiLuzio:2017vat,Calibbi:2017qbu}.
 In this paper we follow the latter direction. 
  
Ultraviolet completions for the vector LQ mediator $U_\mu$ naturally point toward variations of the 
Pati-Salam (PS) gauge group, PS=$\mathrm{SU(4)}\times \mathrm{SU(2)_L }\times \mathrm{SU(2)_R}$~\cite{Pati:1974yy}, 
that contains a massive gauge field with these 
quantum numbers. 
The original PS model does not 
work since the (flavor-blind) LQ field has to be very heavy in order to satisfy the tight bounds from the coupling to the light generations. 
An interesting proposal to overcome this problem has been put forward in Ref.~\cite{DiLuzio:2017vat},
with an extension of  the PS gauge group
and the introduction of heavy vector-like fermions, such that the LQ boson couples to SM fermions only as a result 
of a  specific mass mixing between exotic and SM fermions.

A weakness of most of the explicit SM extensions proposed so far to address the $B$-physics anomalies,
including the proposal of Ref.~\cite{DiLuzio:2017vat},
is the fact that the flavor structure of the  models is somehow ad hoc. This should be contrasted with the 
EFT solution of Ref.~\cite{Buttazzo:2017ixm}, which seems to point toward a common origin between flavor anomalies 
and the hierarchies of the SM Yukawa couplings. In this paper we try to address these problems together, 
proposing a model that is not only able to address the anomalies, but is also able to explain in a natural 
way the observed flavor hierarchies. 

The model we propose is a three-site version of the original PS model. At high energies, the 
gauge group is $\mathrm{PS^3}\equiv\mathrm{PS_1}\times\mathrm{PS_2}\times\mathrm{PS_3}$, where each PS group acts on a single fermion family.
The spontaneous symmetry breaking (SSB) down to the SM group
occurs in a series of steps characterized by different energy scales, which allow us to decouple 
the heavy exotic fields coupled to the first two generations at very high energies. As a result, 
the gauge group controlling TeV-scale dynamics contains a LQ field that is coupled mainly to the 
third generation (see Fig.~\ref{fig:SSB}).  A key aspect of this construction is the hypothesis that 
electroweak symmetry breaking (EWSB) occurs via a Higgs field sitting only on the third-generation site: 
this assumption allows us to derive the hierarchical structure 
of the Yukawa couplings as a consequence of the hierarchies of the vacuum expectation values (VEVs) 
controlling the breaking of the initial  gauge group down to the SM.
In particular, the $\mathrm{U(2)_Q}\times \mathrm{U(2)_L}$ global flavor symmetry appears
as a subgroup of an approximate flavor symmetry of the system emerging at low energies  [$\mathrm{U(2)^5}$].
Last but not least, the localization of the Higgs field on the third-generation site provides a natural screening 
mechanism for the Higgs mass term against the heavy energy scales related to the symmetry breaking 
of the heavy fields coupled to the light generations.

\section{The model}\label{sec:model}

The gauge symmetry of the model holding at high energies 
is $\mathrm{PS}^3\equiv\mathrm{PS}_1\times\mathrm{PS}_2\times\mathrm{PS}_3$, where
\be
\mathrm{PS}_i = \mathrm{SU(4)}_i\times \mathrm{[SU(2)_L]}_i\times \mathrm{[SU(2)_R]}_i~.
\ee
The fermion content is the same as in the SM plus three right-handed neutrinos, such that each fermion family
is embedded in left- and right-handed multiplets of a given $\mathrm{PS}_i$ subgroup:
\be
\Psi_L^\s{i}\sim(\mathbf{4},\mathbf{2},\mathbf{1})_i\,, \qquad 
\Psi_R^\s{i}\sim(\mathbf{4},\mathbf{1},\mathbf{2})_i\,.  \label{eq:SU4psi}
\ee
The subindex $i=1,2,3$ denotes the site that, before any symmetry breaking, can be identified with 
the generation index. 

\begin{figure}[t]
\includegraphics[width=0.42\textwidth]{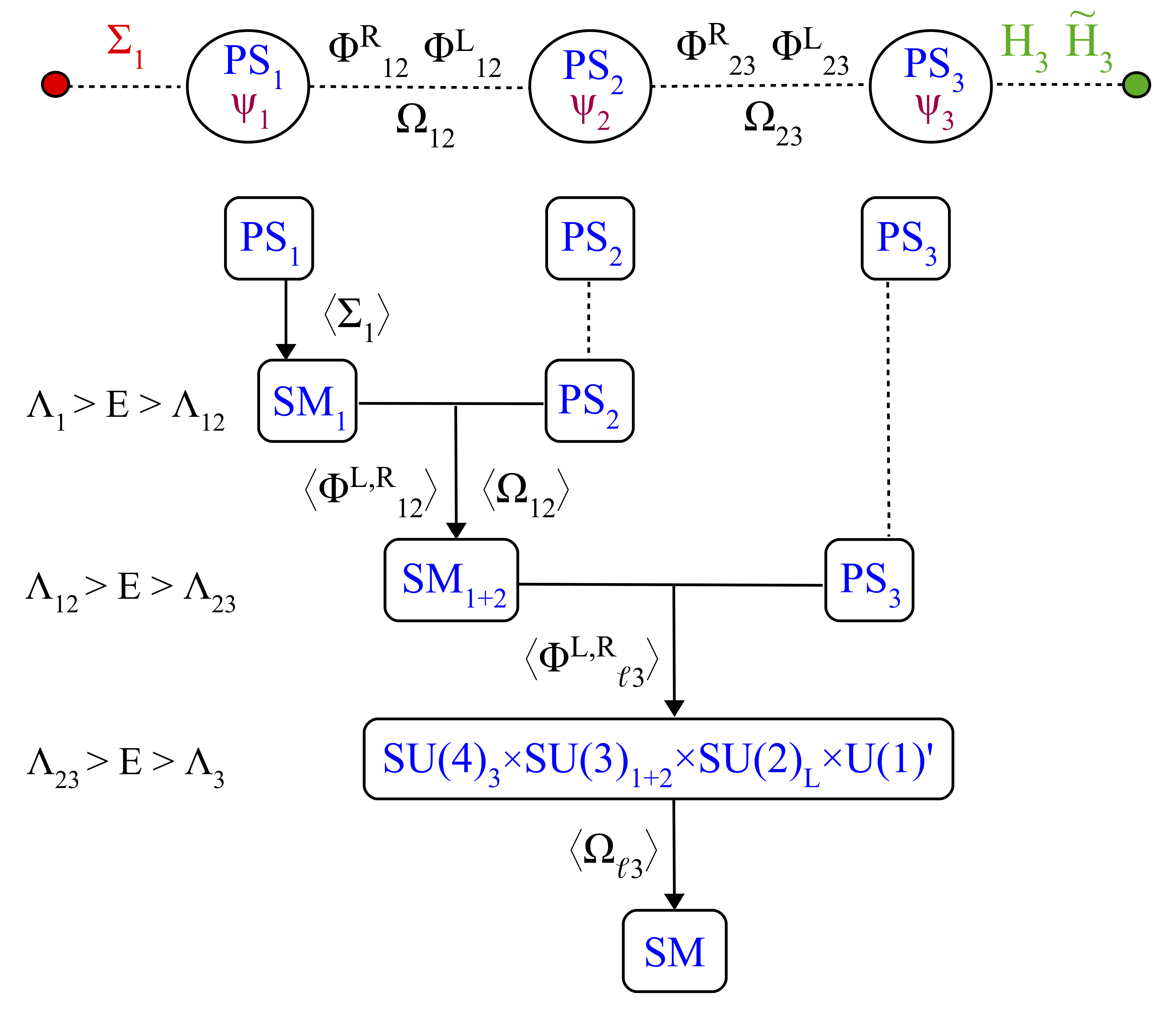}
\caption{Moose diagram of the model (up) and symmetry breaking sequence.}
\label{fig:SSB}
\end{figure}

The SM gauge group is a subgroup of the diagonal group,  $\mathrm{PS}_{\rm diag} = \mathrm{PS}_{1+2+3}$, 
which corresponds to the original PS gauge group. The SSB breaking 
$\mathrm{PS}^3 \to {\rm SM}$ occurs in a series of steps at different energy scales
(see  Fig.~\ref{fig:SSB})
with appropriate scalar fields acquiring non-vanishing VEVs,
as described below.

\medskip
{\bf I.}  High-scale {\em vertical breaking}  [$\mathrm{PS}_1 \to \mathrm{SM}_1$].\\
At some heavy scale, $\Lambda_1  > 10^3$~TeV, the $\mathrm{PS}_1$ group is broken  to 
$\mathrm{SM}_1$, where
\be
\mathrm{SM}_i = \mathrm{SU(3)}_i\times \mathrm{[SU(2)_L]}_i\times \mathrm{[U(1)_Y]}_i~,
\ee
by the VEV of a scalar field $\Sigma_1\sim(\mathbf{4},\mathbf{1},\mathbf{2})_1$,
charged only under  $\mathrm{PS}_1$ (or localized on the first site). Via this breaking 
9 gauge fields with exotic quantum numbers (6 LQ fields, a $W_R^\pm$, and a $Z'$, all coupled only to the first generation) 
acquire a heavy mass and decouple.

\medskip
{\bf II.}   {\em Horizontal breaking} 1--2  [$ \mathrm{SM}_1 \times \mathrm{PS}_2 \to \mathrm{SM}_{1+2}$].\\
Gauge fields on different sites are broken to their diagonal subgroup via appropriate {\em link} fields, or 
scalar bilinears. On both links (1--2 and 2--3) we introduce the following set of link fields
\begin{align}
\label{eq:linkfields}
\begin{aligned}
\Phi_{ij}^L&\sim(\mathbf{1},\mathbf{2},\mathbf{1})_i\times(\mathbf{1},\mathbf{\bar 2},\mathbf{1})_j~, \\
\Phi_{ij}^R&\sim(\mathbf{1},\mathbf{1},\mathbf{2})_i\times(\mathbf{1},\mathbf{1},\mathbf{\bar 2})_j~,\\
\Omega_{ij}&\sim(\mathbf{4},\mathbf{2},\mathbf{1})_i\times(\mathbf{\bar 4} ,\mathbf{\bar 2},\mathbf{1})_j~,
\end{aligned}
\end{align}
such that 
\begin{align}
\begin{aligned}
\langle \Phi_{ij}^{L} \rangle \not = 0   \,  \Rightarrow &  \,
\mathrm{[SU(2)_{L}]}_i\times \mathrm{[SU(2)_{L}]}_j \to \mathrm{[SU(2)_{L}]}_{i+j}~,   \\
\langle \Phi_{ij}^{R} \rangle \not = 0   \, \Rightarrow & \,
\mathrm{[SU(2)_{R}]}_i\times \mathrm{[SU(2)_{R}]}_j \to \mathrm{[SU(2)_{R}]}_{i+j}~,   \\
\langle \Omega_{ij} \rangle \not = 0  \, \Rightarrow &  \, 
\left\{ \begin{array}{l} 
\mathrm{SU(4)}_i \times \mathrm{SU(4)}_j   \to \mathrm{SU(4)}_{i+j} \\
\mathrm{[SU(2)_{L}]}_i\times \mathrm{[SU(2)_{L}]}_j \to \mathrm{[SU(2)_{L}]}_{i+j}~.
\end{array}
\right. \nonumber
\end{aligned}
\end{align}
At a scale $ \Lambda_{12} < \Lambda_1$ the 1--2 link fields acquire a VEV.
As a result, the vertical breaking occurring on the first site is mediated also to the second site,
and the gauge symmetry is reduced to $\mathrm{SM}_{1+2} \times \mathrm{PS}_3$.

Thanks to this second breaking, 9 exotic gauge fields
coupled mainly to the second generation, and 12 SM-like gauge 
fields coupled in a non-universal way to the first two families
acquire a heavy mass and can be integrated out. Below the scale $\Lambda_{12}$ the 
residual dynamical gauge sector is invariant under a global $\mathrm{U(2)^5}$ flavor symmetry acting on the 
first two generations of SM fermions~\cite{nuR}.

At this stage there is still no local coupling between the fermions of the first two generations 
and the scalar fields sitting on the third site ($H_3$ and $\widetilde H_3$) that contain 
the SM Higgs. In other words, we have not yet generated an effective Yukawa coupling for the 
light generations. 

The hierarchy between $ \Lambda_{1}$, $\Lambda_{12}$, and the VEVs of the 1--2
link fields does not need to be specified. The lower bound 
on the lowest of such scales, that we fix to be $10^3$~TeV,
is set by the tight limits on flavor-changing neutral currents 
involving the first two generations (most notably  $K$--$\bar K$ and 
$D$--$\bar D$ mixing \cite{Isidori:2013ez}, and $K_L\to \mu e$~\cite{Giudice:2014tma}). 
With this choice, we can ignore  the effect of  $d\ge 6$ effective operators 
generated at this scale. 

\medskip
{\bf III.}   {\em Horizontal breaking} 2--3  [$\mathrm{SM}_{1+2} \times \mathrm{PS}_3 \to \mathrm{SM}$].\\
The scale characterizing the dynamics of the 2--3 link fieds is $\Lambda_{23} \sim10^2~{\rm TeV}$.
We assume a specific hierarchy among this scale and the VEVs of the link fields:
\be
\Lambda_{23}   >  \langle \Phi_{23}^{L,R} \rangle  >  \langle \Omega_{23} \rangle
\equiv \Lambda_3 \sim 1~{\rm TeV}~.
\ee
This hierarchy is a key ingredient to generate the correct pattern for the 
Yukawa couplings (discussed in detail below) and, at the same time,  address 
the flavor anomalies.\\
At energies $ \langle \Phi_{23}^{L,R} \rangle > E >  \Lambda_3$ we can decouple a $W_L^\pm$, a
$W_R^\pm$, and two $Z'$ fields with mass of $\mathcal{O}$(10 TeV), that are too heavy to be probed at colliders and 
have no impact on flavor physics because of the $\mathrm{U(2)^5}$ flavor symmetry. 

Below $\Lambda_{23}$, the dynamical gauge group is reduced to
\be
\mathcal{G} = \mathrm{SU(4)}_3 \times \mathrm{SU(3)}_{1+2}  \times \mathrm{SU(2)_L} \times \mathrm{ U(1)^\prime}~.
\ee
This symmetry group is structurally similar to the one proposed in~\cite{DiLuzio:2017vat},
but its action on SM fermions is different: with the exception of $\mathrm{SU(2)_L}$, all the other subgroups are flavor 
non-universal.  In particular, the action of $\mathrm{ U(1)^\prime}$ coincides with the SM hypercharge on the first two families and 
with $T^3_R$ on the third family.
The final breaking $\mathcal{G} \to$~SM gives rise to 15 massive gauge bosons 
with mass of $\mathcal{O}$(1 TeV): 6 LQ fields, 8 colorons (i.e.~a color octet), and  a $Z'$. By construction, the LQ is coupled only to the third generation, 
as desired in order to address the flavor anomalies.

\medskip
{\bf IV.}  Low-scale  {\em vertical breaking}~[EWSB].\\
The electroweak symmetry breaking is achieved by an effective 
$\mathrm{SU(2)_L}$ scalar doublet, emerging as a light 
component from the following two set of fields
\be
H_3 \sim(\mathbf{15},\mathbf{2},\mathbf{\bar 2})_3~, \qquad 
\widetilde H_3 \sim(\mathbf{1},\mathbf{2},\mathbf{\bar 2})_3~,
\label{eq:Higgs}
\ee
localized on the third site.

\medskip
In the absence of Yukawa couplings, the full Lagrangian of the proposed model is invariant under the accidental global $[\mathrm{U(1)}_{3B+L}]_i$ symmetries, corresponding to the individual fermion number for each family. The Yukawas explicitly break these symmetries, leaving the diagonal combination $\mathrm{U(1)}_{3B+L}$ unbroken. After the SSB of the PS group to the SM one, this accidental symmetry combines with the $[\mathrm{U(1)}_{B-L}]_i$ generators in $\mathrm{SU(4)}_i$, leaving two unbroken global $\mathrm{U(1)}$ symmetries, $\mathrm{U(1)}_B\!:\,B=X_{3B+L}+1/\sqrt{6}\,T^{15}$ and $\mathrm{U(1)}_L\!:\,L=X_{3B+L}-3/\sqrt{6}\,T^{15}$ (with $T^{15}\equiv T^{15}_1+T^{15}_2+T^{15}_3$). These two symmetries correspond to baryon and lepton numbers and are responsible of keeping the proton stable.

\subsection{Yukawa structure }\label{subsec:yukawas}

The flavor structure observed at low energies emerges as a consequence of the localization of fermions and scalars on different sites.
Given the Higgs fields in \eqref{eq:Higgs}, the only renormalizable (unsuppressed) 
Yukawa interaction at high energies is 
\be
\mathcal{L}_{\rm Yuk}^{\rm ren}=y_3\,\mathrm{Tr}\left\{\overline{\Psi}_L^\s{3}H_3\Psi_R^\s{3}\right\}+\widetilde y_3\,\mathrm{Tr}\left\{\overline{\Psi}_L^\s{3} \widetilde H_3\,\Psi_R^\s{3}\right\}+ {\rm h.c.}
\nonumber 
\ee
and similarly for the conjugate fields  $H_3^c$ and $\widetilde H^c_3$.
The  EWSB breaking induced by $\langle H_3 \rangle$ and $\langle \widetilde H_3\rangle$, 
with $\langle  H_3 \rangle$ aligned along the $T^{15}$ generator of $\mathrm{SU(4)}$,  
allows us to generate four independent SM-like Yukawa couplings for the third generation fermions 
with different SM quantum numbers.

As anticipated, below the scale $\Lambda_{12}$ the dynamical gauge sector is invariant 
under a global $\mathrm{U(2)^5}$ flavor symmetry acting on the 
first two generations of SM fermions:
\begin{align}
\begin{aligned}
\Psi_F^\s{\ell}&\equiv\left(\Psi_F^\s{1},\Psi_F^\s{2}\right)~, \qquad F = \{F_L, F_R\}~,  
\end{aligned}
\end{align}
with $F_L=Q_L,L_L$ and $F_R=U_R,D_R,E_R$.
Effective Yukawa couplings for these fields are generated below the scale $\Lambda_{23}$ (see discussion in Section~\ref{subsec:YukawaOrigin}).
At dimension-five, the following effective operators are generated
\begin{align}\label{eq:d5spurions}
\begin{aligned}
\mathcal{L}_{\rm Yuk}^{d=5}=&\frac{\widetilde y^F_{3\ell}}{\Lambda_{23}}\,\mathrm{Tr}\left\{\overline{\Psi}_{F_L}^\s{\ell}\,\Omega_{\ell3}\, \widetilde H_3\,\Psi_R^\s{3}\right\}+{\rm h.c.}
\end{aligned}
\end{align}
Note that, while the $\mathrm{U(2)^5}$ flavor symmetry is exact in the gauge sector, this is not the case for the scalar sector.
In particular, the $\Omega_{23}$ link field is expected to acquire a non-negligible mixing with $\Omega_{12}$ 
of order $\epsilon_{12} = \langle \Omega_{12} \rangle/\Lambda_{12}  \ll 1$ (and similarly for the other link fields).
This is why we denote $\Omega_{\ell3}$ (rather than $\Omega_{23}$) its dynamical component for $E< \Lambda_{12}$.
Strictly speaking, at this stage we should also treat separately the components of
 $\Omega_{\ell3}$ along the ${\rm SM}_{1+2}$ sub-groups of ${\rm PS}_{1+2}$; 
 however, we leave this tacitly implied.

As  a result of $\mathcal{L}_{\rm Yuk}^{d=5}$, at low energies two spurions of the 
$\mathrm{U(2)_Q} \times \mathrm{U(2)_L} \in \mathrm{U(2)^5}$ flavor symmetry appear.
These spurions (transforming as  $\mathbf{2}_Q$ and $\mathbf{2}_L$, respectively)
control the left-handed mixing between third- and light-generations. Up to $\mathcal{O}(1)$ 
parameters, the size of the  $\mathbf{2}_Q$  spurion can be deduced from the size of the 3--2 mixing in the 
CKM matrix~\cite{Barbieri:2011ci}, implying 
\be
\langle \Omega_{\ell3} \rangle/\Lambda_{23} \sim |V_{ts}| \approx 4 \times 10^{-2}~.
\label{eq:Vts}
\ee
Masses and mixing for the first two generations are obtained from subleading spurions appearing at 
the dimension-six level,
\begin{align}\label{eq:d6spurions}
\begin{aligned}
\mathcal{L}_{\rm Yuk}^{d=6}=&\frac{\widetilde y^F_\ell}{\Lambda_{23}^2}\,\mathrm{Tr}\left\{\overline{\Psi}_{F_L}^\s{\ell}\Phi_{\ell3}^L\, \widetilde H_3 \,\Phi_{3\ell}^R\Psi_{F_R}^\s{\ell}\right\}+{\rm h.c.}
\end{aligned}
\end{align}
Adding these symmetry breaking terms to the ones in \eqref{eq:d5spurions}, we get the following Yukawa pattern
\begin{align}
Y_f =
\begin{pmatrix}
y^f_\ell \frac{\langle\Phi_{\ell3}^L\rangle \langle\Phi_{3\ell}^R\rangle}{\Lambda_{23}^2}& y^f_{3\ell} \frac{\langle\Omega_{\ell3}\rangle}{\Lambda_{23}}\\
0 & y^f_3 
\end{pmatrix}\,,
\label{eq:Yukawa}
\end{align}
where the $y^f_{\ell,3\ell,3}$ are obtained by  $y_{3}$,  $\widetilde y_3$,  and $\widetilde y^F_{\ell,3\ell}$,
normalizing the components of $\langle H_3 \rangle$ and  $\langle \widetilde H_3 \rangle$ to  $v$. 
This structure leads to a very good description of the SM Yukawa couplings 
in terms of $\mathcal{O}(1)$ parameters and VEV ratios. The natural scale for the $d=6$ terms is 
\be
\frac{\langle\Phi_{\ell3}^L\rangle \langle\Phi_{3\ell}^R\rangle}{\Lambda_{23}^2} \sim 
y_c (v) = \frac{m_c(v)}{v}  \approx 5 \times 10^{-3}~.
\ee

A detailed discussion of the scalar sector of the model is beyond the scope of this  
paper. However, it is worth stressing that the various scale hierarchies 
are partially stabilized by the different localization 
of the fields (or by the initial gauge symmetry). In particular, because of \eqref{eq:Vts},
corrections to the Higgs mass term proportional to $\Lambda_{23}^2$ are suppressed 
by $|V_{ts}|^2$, hence they are effectively of $\mathcal{O}$(1~TeV$^2$).

\subsection{Origin of the effective Yukawa operators}
\label{subsec:YukawaOrigin}
The effective Yukawa operators in Section~\ref{subsec:yukawas} cannot be generated 
using only the link fields so far introduced, assuming a renormalizable structure 
at high energies, but can be generated integrating out additional heavy 
fermions or heavy scalar fields with vanishing VEV. In particular, we envisage 
the following three main options:
\begin{itemize}
\item[i)] {\em New link fields}. Adding the following set of (scalar) link fields, 
\begin{align}
\begin{aligned}
\Delta_{ij}\sim\left(\mathbf{4},\mathbf{2},\mathbf{1}\right)_i\times\left(\mathbf{\bar 4},\mathbf{1},\mathbf{\bar 2}\right)_j\,,
\end{aligned}
\end{align}
with vanishing VEV,  we can generate all the effective Yukawa operators at the tree-level via appropriate 
triple and quartic scalar couplings with the other link fields, and (renormalizable) Yukawa-type interactions 
with the chiral fermions. 

\item[ii)] {\em Vector-like fermions}.
The following set of vector-like fermions,
\begin{align}\label{eq:VLcompletion}
\begin{aligned}
\chi_{L/R}&\sim\left(\textbf{4},\textbf{2},\textbf{1}\right)_3\,,\\
\chi_{L/R}^\prime&\sim\left(\mathbf{4},\mathbf{1},\mathbf{1}\right)_i\times\left(\mathbf{1},\mathbf{2},\mathbf{1}\right)_3\,,\\
\chi_{L/R}^{\prime\prime}&\sim\left(\mathbf{4},\mathbf{1},\mathbf{1}\right)_i\times\left(\textbf{1},\textbf{1},\textbf{2}\right)_3\,,
\end{aligned}
\end{align}
is sufficient to induce the desired operators at the tree-level via appropriate new 
Yukawa-type interactions with the link fields and the chiral fermions.  

\item[iii)] {\em Mixed solution}. An interesting mixed solution consists on having a single extra vector-like fermion and a single additional link field, 
\begin{align}
\begin{aligned}
\Delta_{12}&\sim\left(\mathbf{4},\mathbf{2},\mathbf{1}\right)_1\times\left(\mathbf{\bar 4},\mathbf{1},\mathbf{\bar 2}\right)_2\,,\\
\chi_{L/R}&\sim\left(\textbf{4},\textbf{2},\textbf{1}\right)_3\,.
\label{eq:Mixedcompletion}
\end{aligned}
\end{align}
This way the vector-like fermion is responsible of generating the operator in~\eqref{eq:d5spurions},
while the operator in~\eqref{eq:d6spurions} is induced integrating out the new link field.
\end{itemize}

Other possibilities to generate these operators, in particular via loops of extra scalars and fermions, are also possible.
Similarly to the case of the scalar potential, a detailed discussion of the dynamics of these heavy fields 
is beyond the scope of this paper.  On the other hand, it is important discuss in general terms  the nature of 
the higher-dimensional operators, bilinear in the SM fermion fields, generated below the $\Lambda_{23}$ scale
upon integration of generic heavy dynamics. The only two hypotheses we need to assume 
are that: i)~this dynamics respect the $\mathrm{U(2)}^5$ flavor symmetry; ii)~only the link fields in
(\ref{eq:linkfields}) break this symmetry via their VEV.
These two hypotheses are sufficient to ensure a constrained structure for the corresponding EFT,
leading to a well-defined pattern of NP effects at low energies.

The higher dimensional operators can be divided into two main classes:
\begin{itemize}
\item[i)] \textit{$\mathrm{U(2)}$ preserving operators}.  A large set of operators in this category are those containing SM fields only, 
belonging to the so-called SMEFT~\cite{Grzadkowski:2010es}. Other operators  contain $\mathrm{U(2)}^5$-conserving contractions 
of the link fields, or field-strength tensors of the TeV-scale exotic gauge fields. In both cases, the $\mathrm{U(2)}^5$ protection 
and the large effective scale ($\Lambda_{23}\sim 10^2$~TeV) imply marginal effects in 
low-energy phenomenology. 
\item[ii)] \textit{$\mathrm{U(2)}$ breaking operators.} Contrary to the previous case, these operators necessarily involve link fields, namely 
$\Omega_{\ell 3}$, $\Phi_{\ell 3}^L$ and $\Phi_{\ell 3}^R$.  Restricting the attention to the fermion bilinears, it is easy to show that dimension-5 operators 
involve only heavy-light fermions and a single $\Omega_{\ell 3}$ field. These are the Yukawa operators in (\ref{eq:d5spurions}), and operators that reduce to these 
ones after using the equations of motion. 

At dimension six we find operators involving light fermions only and two link fields. The chirally-violating 
ones are the Yukawa terms in~\eqref{eq:d6spurions}. The chirally-preserving ones necessarily involve two powers of the same 
link field. Terms bilinear in $\Phi_{\ell 3}^L$ and $\Phi_{\ell 3}^R$ modify the couplings of the 
heavy $W_L^\pm$, $W_R^\pm$, $Z'$ with mass of $\mathcal{O}$(10 TeV). Given the heavy masses of these fields, 
and the smallness of the $\mathrm{U(2)}$ breaking, these terms are irrelevant for low-energy phenomenology.
We thus conclude that, beside the Yukawa couplings, the only additional effective fermion bilinears 
generated by integrating out heavy dynamics at the scale $\Lambda_{23}$ are operators of the type
\begin{align}
&\frac{i\, \COLL^\s{0}}{\Lambda_{23}^2}{\rm Tr}\{ \Omega_{\ell 3}^\dagger  D^\mu \Omega_{\ell 3}\} (\overline{\Psi}_{F_L}^\s{\ell}  \gamma_\mu \Psi_{F_L}^\s{\ell})~, 
\label{eq:epsilon} \\
&\frac{i\,  \COLL^\s{4}}{\Lambda_{23}^2}{\rm Tr}\{ \Omega_{\ell 3}^\dagger T^\alpha D^\mu \Omega_{\ell 3}\} (\overline{\Psi}_{F_L}^\s{\ell} T_\alpha \gamma_\mu \Psi_{F_L}^\s{\ell})~,
\label{eq:epsilon2}
\end{align}
and analogous terms where $T^\alpha$ is replaced by a $\mathrm{SU(2)_L}$ generator or a combination of $\mathrm{SU(2)_L}$  and $\mathrm{SU(4)}$ generators, 
and finally  terms obtained substituting $\Psi_{F_L}^\s{\ell}$ with $\Psi_{F_R}^\s{\ell}$. 
\end{itemize}

After SSB, the operators (\ref{eq:epsilon})--(\ref{eq:epsilon2}) induce small modifications to the couplings among the TeV-scale gauge bosons and first- and second-generation fermions. 
As we discuss in Section~\ref{sec:pheno}, this effect plays a fundamental role in the explanation of the (subleading) $b\to s\ell\bar\ell$ anomalies. 
On the contrary, the effect of the analogous operators with right-handed fermions are severely constrained by $B_s\to\ell\ell$ ($\ell=e,\mu$). 
It is quite natural to find heavy dynamics that, in first approximation, induces only the left-handed operators and not the right-handed counterparts.
This is for instance the case of  the vector-like fermions in~\eqref{eq:VLcompletion} and \eqref{eq:Mixedcompletion}.
In what follows we include the the operators (\ref{eq:epsilon})--(\ref{eq:epsilon2}) 
in our analysis and neglect the right-handed ones.

\subsection{Gauge boson spectrum at the TeV scale}\label{subsec:TeVmodel}

In what follows we focus on the last step of the breaking chain discussed above, 
namely the $\mathcal{G} \to$~SM breaking, that controls low-energy phenomenology 
and high-$p_T$ physics. We denote the gauge couplings respectively by $g_c^\s{3}$, $\gcl$, $g_L$, and $\gBp$ and the gauge fields by $H_{3\,\mu}^\alpha$, $H_{l\,\mu}^a$, $W_\mu^i$ and $B^\prime$, with $\alpha=1,\dots,15$, $a=1,\dots,8$, and $i=1,2,3$. As discussed above, this symmetry breaking is triggered by the VEV of $\Omega_{\ell3}$, which can be decomposed 
as $\Omega_{\ell3}\stackrel{\mathcal{G}}{\sim}\left(\mathbf{\bar 4},\mathbf{3},\mathbf{3}\right)_{1/6}\oplus\left(\mathbf{\bar 4},\mathbf{1},\mathbf{3}\right)_{-1/2}\oplus\left(\mathbf{\bar 4},\mathbf{3},\mathbf{1}\right)_{1/6}\oplus\left(\mathbf{\bar 4},\mathbf{1},\mathbf{1}\right)_{-1/2}$. We assume that the scalar potential is such that $\Omega_{\ell3}$ only takes a VEV along the $\mathrm{SU(2)_L}$-preserving directions, denoted as $\Omega_3\equiv\left(\mathbf{\bar 4},\mathbf{3},\mathbf{1}\right)_{1/6}$ and $\Omega_1\equiv\left(\mathbf{\bar 4},\mathbf{1},\mathbf{1}\right)_{-1/2}$, while the $\mathrm{SU(2)_L}$-triplet components become heavy and decouple. We have:
\begin{align}
\langle\Omega_{3}\rangle&=\frac{1}{\sqrt{2}}
\begin{pmatrix}
\omega_3 & 0 & 0\\
0 & \omega_3 & 0\\
0 & 0 & \omega_3\\
0 & 0 & 0\\
\end{pmatrix}
\,,\quad
\langle\Omega_{1}\rangle=\frac{1}{\sqrt{2}}
\begin{pmatrix}
0\\
0\\
0\\
\omega_1\\
\end{pmatrix}
\,,
\end{align}
with $\omega_{1,3}$ assumed to be of $\mathcal{O}(\mathrm{TeV})$. These scalar fields can be decomposed under the unbroken SM subgroup as $\Omega_3\sim(\textbf{8},\textbf{1})_0\oplus(\textbf{1},\textbf{1})_0\oplus(\textbf{3},\textbf{1})_{2/3}$ and $\Omega_1\sim(\mathbf{\bar 3},\textbf{1})_{-2/3}\oplus(\textbf{1},\textbf{1})_0$. So, after removing the Goldstones, we end up with a real color octect, one real and one complex singlet, and a complex leptoquark.

The resulting gauge spectrum is the same as in the model proposed in Ref.~\cite{DiLuzio:2017vat}. The massive gauge bosons 
are a vector leptoquark, 
a color octect, and a neutral gauge boson, transforming under the SM subgroup as: $U\sim(\mathbf{3},\mathbf{1})_{2/3}$, $G^\prime\sim(\mathbf{8},\mathbf{1})_0$, and $Z^\prime\sim(\mathbf{1},\mathbf{1})_0$. These are given by the following combinations of the original gauge fields:
\bea
\begin{aligned}
 U_\mu^{1,2,3}&=\frac{1}{\sqrt{2}}\left(H_{3\,\mu}^{9,11,13}-iH_{3\,\mu}^{10,12,14}\right)\,,\\
 G_\mu^{\prime\, a}&=\frac{g_c^\s{l}}{ \gcav } H_{l\, \mu}^a- \frac{g_c^\s{3}}{ \gcav } H_{3\, \mu}^a\,,\\
 Z^\prime_\mu &= \frac{g_c^\s{3}}{ \gBav } H_{3\, \mu}^{15}-\sqrt{\frac{2}{3}} \frac{\gBp}{ \gBav } B^\prime_\mu~,
\end{aligned}
\eea
with  $\gcav= \sqrt{(g_c^\s{3})^2+(g_c^\s{l})^2}$, $\gBav= \sqrt{(g_c^\s{3})^2+\frac{2}{3}(\gBp)^2}$,
and their masses read
\bea
M_{U}&=&\frac{g_c^\s{3}}{2}\sqrt{\omega_1^2+\omega_3^2}\,, \qquad 
M_{G^\prime}=\frac{1}{\sqrt{2}} \gcav \omega_3\,,\nonumber \\
M_{Z^\prime}&=& \frac{3}{2\sqrt{6}} \gBav \sqrt{\omega_1^2+\frac{\omega_3^2}{3}}\,.
\eea
For the phenomenological analysis, it is useful to define the following combination, 
$C_U\equiv v^2\,(g_c^\s{3})^2/4M_{U_3}^2=v^2/(\omega_1^2+\omega_3^2)$, 
which quantifies the overall strength of the NP effects mediated by the vectors at low energies.

The combinations orthogonal to $G_\mu^{\prime\, a}$ and $ Z^\prime_\mu$ are the (massless) 
SM gauge fields $G_\mu^a$ and $ B_\mu$, with couplings 
\be
g_c=\frac{g_c^\s{l}g_c^\s{3}}{ \gcav}~,   \qquad g_Y=  \frac{ \gBp g_c^\s{3}}{ \gBav}~.
\ee
At the matching scale, $\mu\approx 1$~TeV, we have $g_c=1.02$ and $g_Y=0.363$.  From these relations 
it is clear that $g_c^\s{3}, \gcl >g_c$ and $g_c^\s{3}, \gBp >g_Y$, with one of the NP couplings approaching the SM value from above in the limit when the other becomes large. Hence, it follows that $g_c^\s{3}, \gcl \gg \gBp $.
 
A key difference between the model presented here and the one in Ref.~\cite{DiLuzio:2017vat} is found in the couplings of the extra gauge bosons to fermions. In the $\mathrm{SU(4)}$ eigenstate basis (denoted by primed fields) these are given by 
\bea
\mathcal{L}_L&\supset&\frac{g_c^\s{3}}{\sqrt{2}}\,U^\mu\,\overline{q}_L^\prime N^L_U \gamma_\mu\,\ell_L^\prime+{\rm h.c.}\no\\
&&+g_c\,G_\mu^{\prime\,a}\,\overline{q}_L^\prime\, N_{G^\prime} \,\gamma^\mu\,T^a\,q_L^\prime  \no\\
&&+\frac{g_Y}{2\sqrt{6}}\,Z^\prime_\mu\left(3\,\overline{\ell}_L^\prime\, N_{Z^\prime} \,\gamma^\mu\,\ell_L^\prime-\overline{q}_L^\prime\, N_{Z^\prime} \,\gamma^\mu\,q_L^\prime\right)\,, \no\\
\mathcal{L}_R&\supset&\frac{g_c^\s{3}}{\sqrt{2}}\,U^\mu\left(\overline{u}_R^\prime N^R_U \gamma_\mu\,\nu_R^\prime+\overline{d}_R^\prime N^R_U \gamma_\mu\,e_R^\prime\right)+{\rm h.c.}  \no\\
&&+g_c\,G_\mu^{\prime\,a}\left(\overline{u}_R^\prime\,N_{G^\prime}\,\gamma^\mu\,T^a\,u_R^\prime+\overline{d}_R^\prime \, N_{G^\prime} \,\gamma^\mu\,T^a\,d_R^\prime\right) \no\\
&&+\frac{g_Y}{2\sqrt{6}}\,Z^\prime_\mu\left[3\,\overline{\nu}_R^\prime\,N_{Z^\prime}^\s{-} \,\gamma^\mu\,\nu_R^\prime+3\,\overline{e}_R^\prime\,N_{Z^\prime}^\s{+}\,\gamma^\mu\,e_R^\prime\right. \no\\
&&\left.-\,\overline{u}_R^\prime N_{Z^\prime}^\s{+}\gamma^\mu\,u_R^\prime-\overline{d}_R^\prime N_{Z^\prime}^\s{-}\gamma^\mu\,d_R^\prime\right]\,,
\label{eq:gcouplings}
\eea
where we have defined the following matrices in flavor space ($N_c=3\,(1)$ for quarks (leptons))
\begin{align}
\begin{aligned}
N^{L,R}_{U}&=\mathrm{diag}\left(0,0,1\right)\,,\quad\;\, & N_{G^\prime}
&=\mathrm{diag}\left(\frac{\gcl }{g_c^\s{3}},\frac{\gcl }{g_c^\s{3}},-\frac{g_c^\s{3}}{\gcl}\right)\,,\\
N_{Z^\prime}^\s{\pm}&=N_{Z^\prime}\pm\frac{2 \gBp }{3g_c^\s{3}}N_c\,\mathbb{1}\,,& N_{Z^\prime}&=\mathrm{diag}\left(\frac{2 \gBp}{3g_c^\s{3}},\frac{2 \gBp }{3g_c^\s{3}},-\frac{g_c^\s{3}}{ \gBp }\right)\,,
\end{aligned}
\nonumber 
\end{align}
which encode the non-universality of the couplings. The effective operators  in \eqref{eq:epsilon}--\eqref{eq:epsilon2}
generate small additional couplings to the left-handed components of the light families, 
almost aligned to the second generation. This effect is particular relevant for the $U^\mu$ couplings, 
where
\begin{align}
N^L_U\to N^L_U \approx \mathrm{diag}\left( 0,\epsilon,1\right)\, ,
\end{align}
with $\epsilon\equiv -1/2\,\,\COLL^{\s{4}}\,\omega_1\omega_3/\Lambda_{23}^2$, 
while  $N^R_U$ remains unchanged.

For phenomenological applications we need to rewrite these interactions in the fermion mass-eigenstate basis.
This is achieved by rotating the fermion fields with the unitary matrices $V_{f_{L(R)}}$, defined by 
$Y_f=V_{f_L}^\dagger {\rm diag} (Y_f) V_{f_R}$. 
As a result of the Yukawa structure in \eqref{eq:Yukawa}, flavor-mixing terms in the right-handed currents can be 
neglected (the corresponding diagonalization matrices become identity matrices 
in the limit of vanishing light-fermion masses). However,
due to the arbitrariness in the normalization of quark and lepton
fields inside the SU(4) spinors in \eqref{eq:SU4psi}, a freedom remains in the relative phase between
left- and right-handed charged currents. 
Assuming no other sources of CP violation beside the CKM matrix, we restrict this phase ($\theta_{LR}$)
to assume the discrete values $\{ 0,\pi \}$. 

The left-handed flavor rotations can be written as
\begin{align}
\begin{aligned}
q_L^\prime&= V_d\,q_L\equiv
V_d
\begin{pmatrix}
V_{\rm CKM}^\dagger\,u_L\\
d_L
\end{pmatrix}
\,,\\
\ell_L^\prime&= V_e\,\ell_L\equiv
V_e
\begin{pmatrix}
U_{\rm PMNS}^\dagger\,\nu_L\\
e_L
\end{pmatrix}
\,,  
\end{aligned} 
\label{eq:flavrot}
\end{align} 
where $V_{d,e}$ are unitary matrices. As a result of these rotations, flavor-changing terms appear
in the couplings of  $U_3$, $G^\prime$ and $Z^\prime$  to left-handed fermions.
Because of the approximate $\mathrm{U(2)^5}$ flavor symmetry,
we expect both $V_d$ and $V_e$ to be close to the identity matrix; 
for simplicity, we assume them to be real and set to zero the rotations involving the first family:
\begin{align}\label{eq:FlavorRot}
{\footnotesize
\begin{aligned}
V_d=
\begin{pmatrix}
1 & 0 & 0\\
0 & \cos\theta_{bs} & \sin\theta_{bs}\\
0 & -\sin\theta_{bs} & \cos\theta_{bs}\\
\end{pmatrix}
\,,\quad
V_e=
\begin{pmatrix}
1 & 0 & 0\\
0 & \cos\theta_{\tau\mu} & \sin\theta_{\tau\mu}\\
0 & -\sin\theta_{\tau\mu} & \cos\theta_{\tau\mu}\\
\end{pmatrix}
\,.
\end{aligned}
}
\end{align} 
Because of \eqref{eq:Vts}, both $\theta_{bs}$ and $\theta_{\tau\mu}$ are naively  
expected to be of $\mathcal{O}(|V_{ts}|)$. However, in order to avoid the strong bounds 
from $B_s$-mixing, we assume $y^d_{3\ell} / y^d_3 \ll 1$, such that $\theta_{bs}\ll |V_{ts}|$.

\section{Phenomenological analysis}\label{sec:pheno}

\begin{figure}[t]
\vskip - 0.5 true cm 
\includegraphics[width=0.38\textwidth]{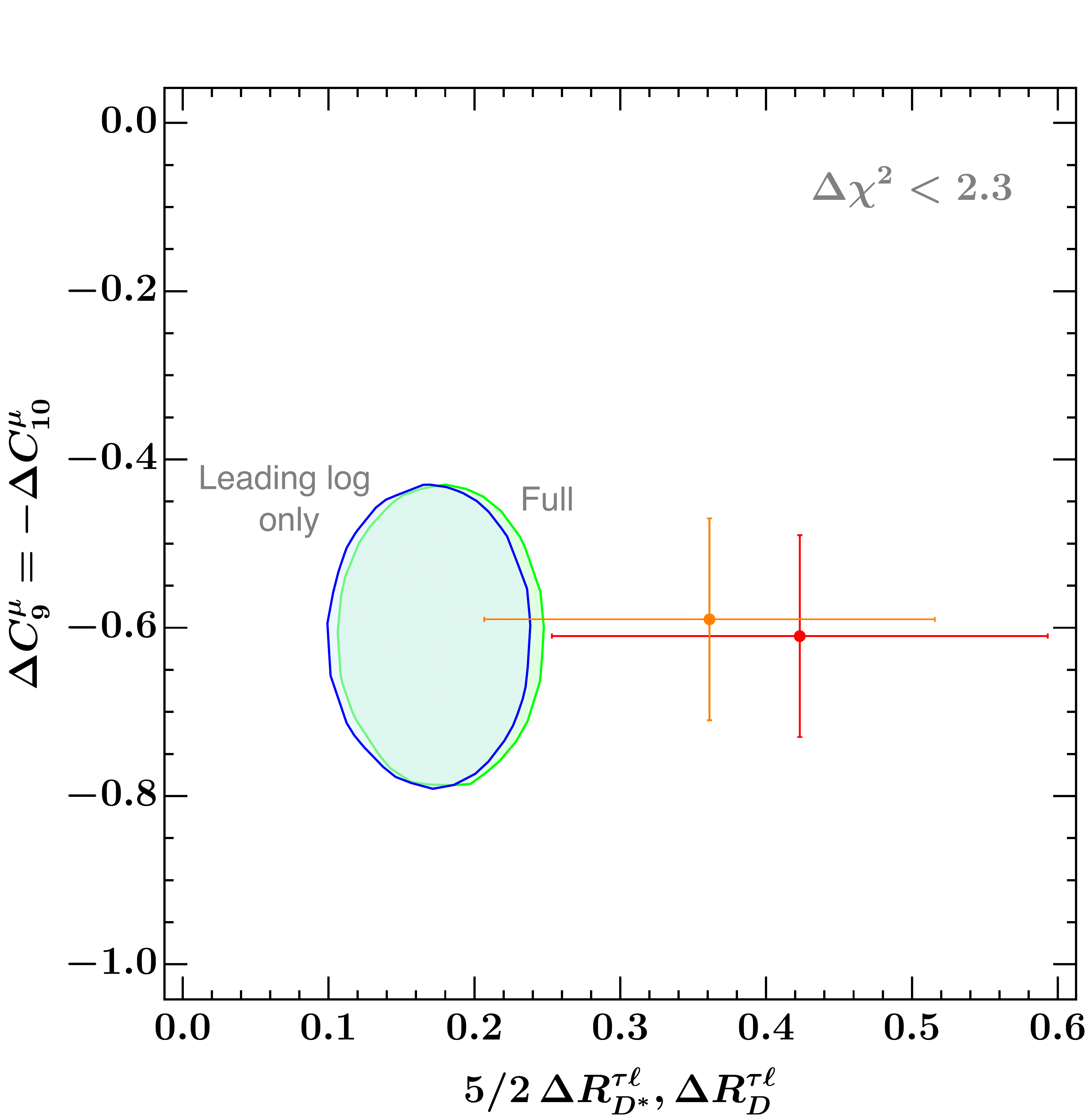}
\caption{Model prediction for $\Delta C_9^\mu=-\Delta C_{10}^\mu$,  $\Delta R_{D^{*}}^{\tau\ell}$, and 
$\Delta R_{D}^{\tau\ell}$
 for the $\Delta\chi^2\leq2.3~(1\sigma)$ fit region: in blue including only the logarithmic contribution in \eqref{eq:radiative}, and in green including also the non-logarithmic corrections. The  $1\sigma$ experimental data are shown by the two crosses.
Predictions and results for $\Delta R_{D^*}^{\tau\ell}$ (red cross) are scaled by $5/2$ compared to $\Delta R_{D}^{\tau\ell}$  (orange cross),
since our model predicts $\Delta R_{D}^{\tau\ell} \approx 5/2 \times  \Delta R_{D^{*}}^{\tau\ell}$.
  }
\label{fig:RDvsC9}
\end{figure}

\textbf{Low-energy constraints.} The low-energy phenomenology of the model can be described in terms of   
 $\{C_U, \epsilon, \theta_{\tau\mu}, \theta_{bs}\}$ and the discrete parameter $\theta_{LR}$.
The list of relevant low-energy observables, with their explicit expression in terms of four-fermion effective operators,
 is given in Table~II of Ref.~\cite{Buttazzo:2017ixm}. 
 An important difference 
 is the appearence of 
effective  charged-current scalar operators from the right-handed terms in (\ref{eq:gcouplings}).
These have a negligible impact in $B\to D^* \tau \nu$, but 
are non-negligible in $B\to D \tau \nu$. Using the results in Ref.~\cite{Fajfer:2012vx} 
for the matrix-elements of the $(\bar b_R c_L) (\bar \nu_L \tau_R)$
operator, we obtain in the limit $\theta_{\tau\mu},\theta_{bs}\to 0$
\bea
\begin{aligned}
	\Delta R_{D^*}^{\tau \ell} &= R_{D^*}^{\tau \ell}  - 1  \approx  2  [ 1- 0.12 \cos(\theta_{LR}) ]C_U~, \\  
	\Delta R_{D}^{\tau \ell}  &= R_{D}^{\tau \ell} -1 \approx  2  [ 1 - 1.5 \cos(\theta_{LR})] C_U~, \qquad 
\end{aligned}
\eea
with $R_{D^{(*)}}^{\tau \ell}$ defined as in Ref.~\cite{Buttazzo:2017ixm}.
In order to maximize the correction  to $R_{D^{(*)}}^{\tau \ell}$
we set $\theta_{LR}=\pi$. This implies the relation 
$\Delta R_{D}^{\tau\ell} \approx 5/2 \times  \Delta R_{D^{*}}^{\tau\ell}$, 
that is well consistent with
present data~\cite{Lees:2013uzd,Aaij:2015yra,Hirose:2016wfn,Aaij:2017deq}.

Having fixed $\theta_{LR}$, we determine the remaining four parameters from a global fit.
At the best fit point we obtain $\chi^2_{\rm min}\approx9$, which gives a very good fit compared to the SM, for which $\chi^2_{\rm SM}\approx46$. A typical set of parameters providing a good fit to data is given by  $C_U=0.03$, $\epsilon=-0.02$, $\theta_{\tau\mu}=-0.05$ and $\theta_{bs}=0.05\,V_{ts}$. This can be obtained for instance from the benchmark point: 
 $g_c^\s{3}=3$ and $M_{U_3}\approx2$~TeV, with $M_{G^\prime}$ and $M_{Z^\prime}$ ranging 
 between 1.5 and 3~TeV (depending on the $\omega_1/\omega_3$ ratio).
 
The potential of the model to explain the anomalies in $b\to s\ell\bar\ell$ 
(that we express as deviations in the Wilson coefficients $C_{9,10}$, 
defined as in~\cite{Altmannshofer:2015sma,Descotes-Genon:2015uva}) 
and in $R_{D^{(*)}}^{\tau\ell}$ (for which we adopt the updated SM prediction in~\cite{Bigi:2017jbd,Jaiswal:2017rve,Bigi:2016mdz})
is depicted in Fig.~\ref{fig:RDvsC9}. A good fit to $b\to s\ell\bar\ell$ data can only be achieved when considering the dimension-six operator \eqref{eq:epsilon}, whose effect is encoded in $\epsilon$. Interestingly, the best fit value for $\epsilon$ is perfectly 
consistent with that of the dimension-six contributions in the Yukawa couplings.

While the model  significantly reduces the tension with data, 
predicting a non-trivial correlation between $R_{D}^{\tau\ell}$ and $R_{D^*}^{\tau\ell}$
 (see caption of Fig.~\ref{fig:RDvsC9}), 
the central value of these two observables cannot be achieved due to the constraints 
from LFU tests in $\tau$ physics and $\mathcal{B}(B_{c,u}\to \tau\nu)$.
The LFU tests yield per-mille constraints on the 
modifications of $W$ and $Z$ couplings to $\tau$ leptons ($\delta g_\tau^W$ and $\delta g^Z_{\tau_L,\nu_\tau}$)~\cite{EWPT}.
These quantities arise in our model from one-loop diagrams involving SM fermions and LQ fields~\cite{Loop}, 
whose (leading) result at $\mathcal{O}(y_t^2)$ is 
\begin{align}\label{eq:radiative}
\begin{aligned}
\delta g_\tau^W/g_\ell^W &=\frac{3\,y_t^2}{16\pi^2}\,C_U\left(\frac{1}{2}+\log \frac{m_t^2}{M_U^2}\right)\,,\\
\delta g^Z_{\nu_\tau}/ g^Z_{\nu_\ell} &=\frac{3\,y_t^2}{8\pi^2}\,C_U\left(1+\log \frac{m_t^2}{M_U^2}\right)\,.
\end{aligned}
\end{align}
These expressions agree in the logarithmic part with the EFT results in~\cite{Jenkins:2013wua,Feruglio:2016gvd,Feruglio:2017rjo}. 
However, having a complete model, we have been able to compute also the non-logarithmic terms which are non-negligible and partially alleviate the tensions with LFU tests in $\tau$ physics (see Fig.~\ref{fig:RDvsC9}). 
As far as $\mathcal{B}(B_{c,u}\to \tau\nu)$ are concerned, at the best fit point we predict 
a $\sim 60\%$ enhancement over the SM, which is perfectly consistent with present data.

Another important constraint is obtained from $B_{s,d}$ mixing. Contributions to these observables arise in our model from the tree-level exchange of the coloron and the $Z^\prime$, as well as from one-loop box diagrams involving the vector leptoquark. All these contributions are proportional  to the down-type rotation angle $|\theta_{bs}|$. Allowing for ($U(2)^5$ preserving) deviations of up to $\mathcal{O}(10\%)$ in  $B_{s,d}$ mixing leads to the bound $|\theta_{bs}|\lesssim0.1\,|V_{ts}|$, forcing a flavor-alignment in the down-quark sector. As a result of this alignment, contributions to $D-\bar D$ mixing from coloron and $Z^\prime$ exchange turn out to be 
below the present limits and do not give any relevant bound.

The vector leptoquark does not contribute significantly to $B\to K^{(*)}\nu\bar\nu$ nor to $\tau\to3\mu$, while the approximate down-alignment in the quark sector required from $B_s$ mixing renders the $Z^\prime$ contribution to $B\to K^{(*)}\nu\bar\nu$ negligibly small. The $Z^\prime$ contributes at tree-level to $\tau\to3\mu$. However, since its coupling to muons is suppressed, the constraints from these processes only become relevant when the leptonic mixing angle $\theta_{\tau\mu}$ becomes large, effectively setting the bound 
$|\theta_{\tau\mu}| \lesssim0.1$. 

\medskip
\textbf{High-$p_T$ searches.} The masses of the lightest exotic vector bosons predicted by the model are expected to lie around the TeV scale, and are therefore constrained by direct searches at LHC. The phenomenology for these searches is very similar to the one discussed in the model of Ref.~\cite{DiLuzio:2017vat}, so we only highlight the main aspects.

\smallskip
\noindent$\bullet$ \textbf{$U$.} The vector LQ is subject to the bounds coming from QCD pair production and from tau pair production at high-energies (i.e. $pp\to\tau\bar\tau+X$), generated by $t$--channel exchange~\cite{Faroughy:2016osc}. As in Ref.~\cite{DiLuzio:2017vat}, the most stringent constraint is set by leptoquark pair production, which implies $M_U\gtrsim1.3$~TeV. This expression is obtained by recasting~\cite{DiLuzio:2017chi} the CMS search in Ref.~\cite{Sirunyan:2017yrk} and translates to $C_U\lesssim0.08$ for $g_c^\s{3}=3$.

\smallskip
\noindent $\bullet$ \textbf{$G^\prime$.}  Given the large couplings and relatively low mass of the coloron, di-jet searches at LHC can offer an important test of the validity of the model. However, current limits~\cite{Aaboud:2017yvp} rely on bump searches that become less sensitive when the coloron width is large. This is the case in our model, where we find $\Gamma_{G^\prime}/M_{G^\prime}=0.22$ for $g_c^\s{3}=3$, if we assume that the only available decay channels are those to SM quarks. For large widths, the coloron signal is diluted into the QCD background allowing the model to avoid current bounds~\cite{Admir:implications}.

\smallskip
\noindent $\bullet$ \textbf{$Z^\prime$.} As already mentioned, the $Z^\prime$ couplings to light generations appear strongly suppressed compared to the third-generation ones. This renders the $Z^\prime$ Drell-Yan production at LHC sufficiently small to evade the strong bounds from di-lepton resonance searches~\cite{Aaboud:2017buh}.

\smallskip
\noindent $\bullet$ \textbf{Heavy scalars.} The minimal model discussed in Section~\ref{subsec:TeVmodel} presents a rich scalar sector, whose phenomenological analysis depends significantly on the details of the scalar potential and is beyond the scope of the present letter. Nevertheless, we do not expect it to yield tensions with data in large areas of the parameter space.

\section{Summary and conclusions}\label{sec:conclusions}
If unambiguously confirmed as beyond-the-SM signals, 
the recent $B$-physics anomalies would lead to a significant 
shift in our understanding of fundamental interactions.
They could imply abandoning the assumption of flavor universality of gauge interactions, 
which implicitly holds in the SM and in its most popular extensions. 
In this paper we have presented a model where the idea of 
flavor non-universal gauge interactions is pushed to its 
extreme consequences, with an independent gauge group 
for each fermion family.

The idea of the (flavor-blind) SM gauge group being the result of a suitable breaking of a flavor non-universal 
gauge symmetry, holding at high energies, has already 
been proposed in the past as a possible explanation for 
the observed flavor hierarchies (see e.g.~\cite{Craig:2011yk,Barbieri:2011ci}). Interestingly, constructions 
of this type naturally arise in higher-dimensional models (see e.g.~\cite{Honecker:2003vq})
with fermion fields localized on different four-dimensional branes, the multi-site gauge group
being the deconstructed version of a single higher-dimensional gauge symmetry~\cite{ArkaniHamed:2001nc}.

As we have shown in this paper,  a three-site Pati-Salam gauge symmetry, with a suitable 
symmetry breaking sector, could describe in a natural way the observed Yukawa hierarchies and explain at the same time the recent $B$-physics anomalies, while being consistent with the tight 
constraints from other low- and high-energy measurements.
The model we present exhibits a rich TeV-scale phenomenology that can be probed in the near future 
by high-$p_T$ experiments at the LHC.

\acknowledgements
We thank R. Barbieri, D. Buttazzo, A. Greljo, D. Marzocca, M. Nardecchia and A. Pattori for useful comments and discussions. J.F. thanks the CERN Theory Department for hospitality while part of this work was performed.
This research was supported in part by the Swiss National Science Foundation (SNF) under contract 200021-159720.


\begin{thebibliography}{99}
\bibitem{nuR}
At $E < \Lambda_{12}$ mass terms for the right-handed neutrinos 
of the first two generations are allowed. We thus integrate out also $\nu^R_{1,2}$ remaining with 
5 independent species of massless fermions charged under $\mathrm{SM}_{1+2}$.
\bibitem{EWPT}
The dominant constrain arises from the bound on $|g_\tau^W/g_\mu^W|$, for which we use the value reported in
  Y.~Amhis {\it et al.},
  [arXiv:1612.07233].
\bibitem{Loop}
An additional contribution to $\delta g^Z_{\tau_L,\nu_\tau}$ arises in our model from the mixing of the lowest-lying $Z^\prime$ and the SM $Z$; however, this receives a parametric suppression of $\mathcal{O}( s_W^2\, \gBp /g_c^\s{3}\,M_Z^2/M_{Z^\prime}^2)$ 
and turns out to be negligible.
\end{thebibliography}
\end{document}